\pdfoutput=1
\documentclass[11pt]{article}

\usepackage[preprint]{acl}

\usepackage{times}
\usepackage{latexsym}
\usepackage{longtable}

\usepackage[T1]{fontenc}
\usepackage[utf8]{inputenc}

\usepackage{microtype}

\usepackage{inconsolata}

\usepackage{graphicx}
\usepackage{lipsum} 
\usepackage{xspace}
\usepackage{enumitem}
\usepackage{microtype}      % microtypography
\usepackage{xcolor}         % colors
\usepackage{floatrow}
\usepackage{multirow}
\usepackage{bm}
\usepackage{amsmath}
\usepackage{multicol}
\usepackage{makecell}
\usepackage{tabularx}
\usepackage[rightcaption]{sidecap}
\usepackage{subcaption}
\usepackage{listings}
\usepackage{booktabs}
\usepackage{wrapfig}
\usepackage{array} % 如果需要自定义列宽，可以使用这个宏包
\usepackage{graphicx}
\usepackage{multirow}
\usepackage{tcolorbox}
\tcbuselibrary{listings, breakable, skins, theorems}
\usepackage[utf8]{inputenc} % allow utf-8 input
\usepackage[T1]{fontenc}    % use 8-bit T1 fonts
\usepackage{hyperref}       % hyperlinks
\usepackage{url}            % simple URL typesetting
\usepackage{booktabs}       % professional-quality tables
\usepackage{amsfonts}       % blackboard math symbols
\usepackage{nicefrac}       % compact symbols for 1/2, etc.
\usepackage{microtype}      % microtypography
\usepackage{xcolor}         % colors
\usepackage{amssymb}
\usepackage{float}
\usepackage{amsmath}
\usepackage{multicol}
\usepackage{makecell}
\usepackage{tabularx}
\usepackage[rightcaption]{sidecap}
\usepackage{subcaption}
\usepackage{multirow}
\usepackage{bm}
\usepackage[utf8]{inputenc} % allow utf-8 input
\usepackage[T1]{fontenc}    % use 8-bit T1 fonts
\usepackage{hyperref}       % hyperlinks
\usepackage{url}            % simple URL typesetting
\usepackage{booktabs}       % professional-quality tables
\usepackage{amsfonts}       % blackboard math symbols
\usepackage{nicefrac}       % compact symbols for 1/2, etc.
\usepackage{microtype}      % microtypography
\usepackage{xcolor}         % colors
\usepackage{amssymb}
\usepackage{listings}
\newcommand{\name}{\textsc{CoIR}\xspace}

\title{
  \begin{minipage}{0.70\textwidth}
    \raisebox{-0.4\height}[0pt][0pt]{
      \includegraphics[height=1.9\baselineskip]{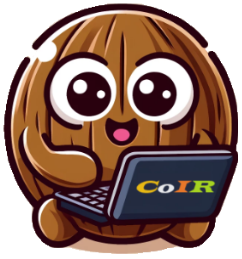}}
      \hspace{-10pt}
    \begin{minipage}{0.85\textwidth}
    \centering
      \name: A Comprehensive Benchmark for \\ Code Information Retrieval Models
    \end{minipage}
  \end{minipage}
}

\author{%
  Xiangyang Li\thanks{These authors contributed equally to this work.} \quad  Kuicai Dong\footnotemark[1] \quad Yi Quan Lee\footnotemark[1] \quad Wei Xia  \quad \\
  \textbf{Hao Zhang}\thanks{Corresponding authors.} \quad \textbf{Xinyi Dai} \quad  \quad \textbf{Yasheng Wang} \quad \textbf{Ruiming Tang}\footnotemark[2] \\
  \textsc{Huawei Noah's Ark Lab} \\
  \texttt{\{lixiangyang34, kuicai.dong, lee.yi.quan, xiawei24, } \\ 
  \texttt{zhang.hao3, daixinyi5,wangyasheng, tangruiming\}@huawei.com}
}

\begin{document}
\maketitle
\begin{abstract}

% Despite the substantial success of Information Retrieval (IR) in various NLP tasks, most IR systems predominantly handle queries and corpora in natural language, rather than code retrieval. 
% Code retrieval is critically important yet remains under-explored, with existing methods and benchmarks inadequately representing the diversity of code in various domains and tasks. Addressing this gap, we present \textbf{\name} (\textbf{Co}de \textbf{I}nformation \textbf{R}etrieval Benchmark), a robust and comprehensive benchmark specifically designed to evaluate code retrieval capabilities. \name comprises \textbf{ten} meticulously curated code datasets, spanning \textbf{eight} distinctive retrieval tasks across \textbf{seven} diverse domains. 
% We first discuss the construction of \name and its diverse dataset composition. 
% Further, we evaluate ten widely used retrieval models using \name, uncovering significant difficulties in performing code retrieval.
% % tasks even with state-of-the-art systems. 
% \name also introduces a simple yet effective python framework.
% % , which additionally defines various advanced modes to facilitate researchers in evaluating their models. 
% It shares same data schema as other popular benchmarks like MTEB and BEIR, enabling seamless cross-benchmark evaluations. Through \name, we aim to invigorate research in the code retrieval domain, providing a versatile benchmarking tool that encourages further development and exploration of code retrieval systems\footnote{\url{ https://anonymous.4open.science/r/coir-8FE4/}}.
Despite the substantial success of Information Retrieval (IR) in various NLP tasks, most IR systems predominantly handle queries and corpora in natural language, neglecting the domain of code retrieval. Code retrieval is critically important yet remains under-explored, with existing methods and benchmarks inadequately representing the diversity of code in various domains and tasks. Moreover, many models have begun to overfit existing leaderboards, limiting their generalizability and real-world applicability. Addressing this gap, we introduce \textbf{\name} (\textbf{Co}de \textbf{I}nformation \textbf{R}etrieval Benchmark), a robust and comprehensive benchmark specifically designed to evaluate code retrieval capabilities. \name consists of \textbf{ten} meticulously curated code datasets, all of which have undergone thorough manual inspection and processing. These datasets cover \textbf{eight} distinct retrieval tasks across \textbf{seven} diverse domains, ensuring a broad and rigorous assessment of code retrieval performance.
We first discuss the construction of \name and its diverse dataset composition. 
Further, we evaluate ten widely used retrieval models using \name, uncovering significant difficulties in performing code retrieval tasks even with state-of-the-art systems. 
To ensure seamless integration, \name is released as a user-friendly Python framework, aligned with the data schema of MTEB and BEIR for consistent cross-benchmark evaluation. Through \name, we aim to invigorate research in the code retrieval domain, providing a versatile benchmarking tool that encourages further development and exploration of code retrieval systems\footnote{\url{ https://github.com/CoIR-team/coir}}.

\end{abstract}

\section{Introduction}\label{sec:intro}

Information retrieval (IR) aims to retrieve
% is a process that facilitates effective and efficient retrieval of 
relevant information from a large scale corpus. The advances of pretrained Transformers~\citep{DBLP:conf/nips/VaswaniSPUJGKP17} like BERT~\citep{devlin-etal-2019-bert} and  RoBERTa~\citep{liu2019roberta} enable IR systems~\citep{wang2022-e5,chen2024bge,li2023-gte,dong2025mmdocir} to leverage rich semantic embeddings to interpret and fulfill natural language search queries. 
Specifically, IR have successfully improve a wide range of Natural Language Processing (NLP) tasks such as Question Answering (QA)~\citep{DBLP:journals/isci/KolomiyetsM11},
% DBLP:journals/widm/AbbasiantaebM21
Retrieval-Augmented Generation (RAG)~\citep{ctrla, mc, dong2025mmdocrag,reasonrag,jia2024bridging}, 
% DBLP:conf/nips/LewisPPPKGKLYR020, DBLP:journals/corr/abs-2404-13781,
Information Extraction~\citep{DBLP:conf/fedcsis/Ziembinski15, DBLP:phd/basesearch/Sarhan23},
% milward-thomas-2000-information,
Text Summarization~\citep{DBLP:journals/access/MahalakshmiF22}, Recommender System~\cite{inttower,ctrl,can,flip}
and etc. %DBLP:journals/eswa/GlavasS14,
Although proven to be effective in text retrieval, standard IR methods often fall short in 
% specialized domains such as 
code retrieval~\citep{husain2019codesearchnet}.

\setlength{\tabcolsep}{1pt} 
\begin{table}[]{}{} % Adjust width
  \resizebox{1\textwidth}{!}{ % Resize table
\begin{tabular}{@{}ccccc@{}}
\toprule
Benchmark   & Domain  & \#PL & \begin{tabular}[c]{@{}c@{}} Retrieval \\ Tasks \end{tabular}  & \begin{tabular}[c]{@{}c@{}}Eval\\ Package\end{tabular} \\ \midrule

CoSQA  & Web Query  & 1   & Text-to-Code  & $\times$   \\ \midrule

CodeSearchNet & GitHub Functions   & 6    & Text-to-Code & $\times$    \\ \midrule

CodeRAG-Bench  & \begin{tabular}[c]{@{}c@{}}Contest, Issue Fixing, \\ StackOverflow, \\ GitHub Functions\end{tabular}  & 1   & \begin{tabular}[c]{@{}c@{}}Text-to-Code \\ Hybrid Code\end{tabular}  & $\times$   \\ \midrule

XCodeEval    & Contest   & 17     & \begin{tabular}[c]{@{}c@{}}Text-to-Code \\ Code-to-Code \end{tabular}   & $\times$   \\ \midrule

\name (\textbf{Ours})  & \begin{tabular}[c]{@{}c@{}} GitHub Functions,\\ Web Query,\\ Database, Contest,\\ Deep Learning, \\ StackOverflow, \\ Code Instruction \end{tabular} & 14  & \begin{tabular}[c]{@{}c@{}}Text-to-Code \\ Code-to-Code\\ Code-to-Text \\ Hybrid Code \end{tabular} & $\checkmark$ \\ \bottomrule
\end{tabular}
  }
  \vspace{-0.8em}
  \caption{Comparison between CoIR and other code retrieval benchmarks. \#PL is the number of main programming languages used.
  \vspace{-1.0em}
  \label{tab:compare}}
\end{table}

Unlike standard text, code is semi-structured and inherently logical, consisting of syntactic rules and semantic information that require specific parsing and understanding. Such distinctive nature requires the system to adapt and interpret code format accurately. Recognizing the importance of code data, pioneering works such as CodeBERT~\citep{feng-etal-2020-codebert}, CodeGPT~\citep{DBLP:conf/nips/LuGRHSBCDJTLZSZ21} and UniXcoder~\citep{guo-etal-2022-unixcoder} have conducted pre-training specifically on code corpora.
% to perform well on code-related tasks. 

Code information retrieval is a critical component in accelerating development processes and improving code quality. Efficient code retrieval helps developers quickly find not only relevant code snippets, but also related information like code explanations, bug analyses, code summaries, and similar code instances. Commercial products have recently integrated tools for code retrieval, such as VS Code~\citep{code2019visual} and GitHub Code Search.
Moreover, code-RAG systems~\citep{zhang-etal-2023-syntax, zhang-etal-2023-repocoder, DBLP:journals/access/ChoiNKL23, DBLP:journals/corr/abs-2402-12317} have effectively leveraged on code retrieval to minimize hallucinations (i.e., errors in generated code) by Large Language Models (LLMs), thereby ensuring more accurate and reliable outputs.

Due to the importance of code retrieval, benchmarks (see Table~\ref{tab:compare}) such as CodeSearchNet~\citep{husain2019codesearchnet}, CosQA~\citep{huang-etal-2021-cosqa}, and XcodeEval~\citep{khan2023xcodeeval}, have been proposed to evaluate the code retrieval effectiveness.
Despite these efforts, there remain three principal limitations: 
% that need to be addressed.
\textbf{(1)} \textit{Current benchmarks focus on a limited number of code retrieval tasks and have been extensively overfitted by many existing models}. Commonly, these involve using a textual query to search for corresponding code snippets. However, the practical needs of code retrieval are far more diverse. In real-world scenarios, queries and retrieved corpus can involve both text and code.
% not just text or code, but a combination of both. 
% For instance, o
One might input a code snippet coupled with bug information, and seek detailed explanations, summaries, or even fixed code as output. Existing benchmarks do not adequately cater to such complex and varied query types, limiting the scope of their applicability and the robustness of the models tested.
\textbf{(2)} \textit{there is a noticeable lack of diversity in data domains}. For example, CodeSearchNet exclusively extracts code and code-comment pairs from GitHub, which represents a specific format of open-source projects.
% of coding practices in open-source projects. 
Similarly, XcodeEval focuses only on coding related to contest challenges, which may be overly specialized. 
Such narrow focus is not suitable for comprehensive evaluation in broader coding contexts.
% Such narrow focus implies that models trained and evaluated on these datasets could exhibit suboptimal performance in broader coding contexts.
\textbf{(3)} \textit{there is no standard evaluation framework for code retrieval}, which complicates the comparison and development of methods in this field. CodeSearchNet, CosQA, and XcodeEval employ various types of evaluation metrics tailored to their specific tasks and formats
% This variety in evaluation approaches means that metrics used in one benchmark are often not directly applicable to others, 
leading to potential inconsistencies in measuring model performance across different benchmarks.

To address the limitations of existing code retrieval benchmarks, we introduce \textbf{\name} (\textbf{\textsc{Co}}de \textbf{\textsc{I}}nformation \textbf{\textsc{R}}etrieval Benchmark), a more comprehensive and versatile benchmark. As summarized in Table~\ref{tab:compare}, \name surpasses current benchmarks by offering broader domain coverage, more diverse retrieval tasks, and a standardized evaluation framework. It includes 10 datasets (8 existing and 2 newly-curated) and supports 4 primary retrieval tasks: (1) Text-to-Code, (2) Code-to-Code, (3) Code-to-Text, and (4) Hybrid Code Retrieval, with further breakdown into 8 sub-tasks. The datasets vary in size, ranging from 1K to 1M documents, with token counts ranging from 37 to 4.4K for queries and 113 to 1.5K for corpus.

We evaluated 10 popular retrieval models on \name, revealing that even state-of-the-art models perform suboptimally, highlighting the complexity of code retrieval. To simplify evaluation, we also provide a user-friendly Python framework that integrates seamlessly with BEIR and MTEB, allowing easy model evaluation via pip installation. Our key contributions are as follows:
\begin{itemize}[left=0pt, itemindent=0pt,itemsep=-1pt]
    \item We present the first benchmark designed to comprehensively evaluate code retrieval. This benchmark, \name, integrates 10 datasets and addresses 4 key code retrieval tasks. To achieve this, we manually collected and curated existing datasets through a rigorous data cleaning and filtering process, while also creating two new datasets.
\item Our evaluation of 10 popular retrieval systems reveals that even state-of-the-art models struggle in code retrieval. Through experimental analysis, we demonstrate that LLM-based retrieval models have the potential to pave the way for new directions in code and text retrieval. Additionally, our findings indicate that many existing models have already overfitted to current leaderboards, highlighting the necessity for more generalizable and robust retrieval approaches. 
    \item We offer a robust community around \name, providing convenient evaluation tools, leaderboards, and other resources that facilitate the rapid advancement of code retrieval models.
\end{itemize}

\begin{figure*}[t] 
    \centering
    \includegraphics[width=0.9\textwidth, trim={0.0cm 0.15cm 0.0cm 0.38cm}, clip]{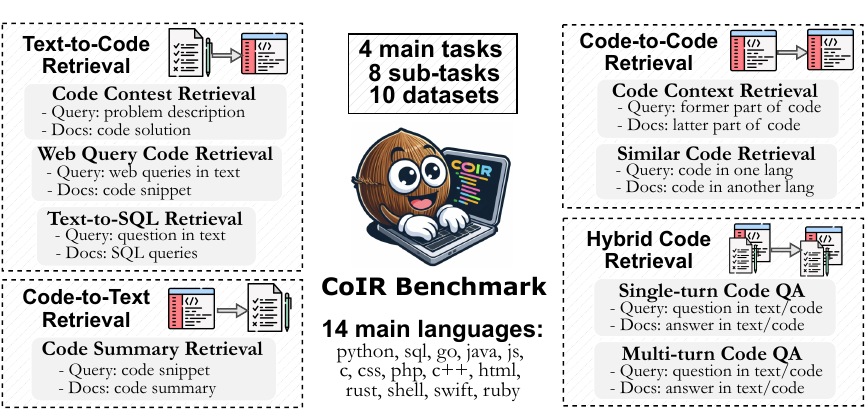} 
    \vspace{+0.8em}
    \caption{Overview of \name benchmark.} 
    \label{fig:coir_overview} 
    \vspace{-0.8em}
\end{figure*}

\section{Related Work}
\label{sec:related}
% \vspace{-0.4em}

\textbf{Existing Benchmarks.}
BEIR~\citep{thakur2021beir} and MTEB~\citep{muennighoff-etal-2023-mteb} have significantly advanced retrieval and embedding models across diverse text-based tasks but lack dedicated evaluations for code retrieval. Existing code benchmarks, such as CodeSearchNet~\citep{husain2019codesearchnet}, CosQA~\citep{huang-etal-2021-cosqa}, and XcodeEval~\citep{khan2023xcodeeval}, suffer from limitations in task diversity, domain coverage, and evaluation consistency. Notably, CodeSearchNet has been extensively used, leading to overfitting in many models. In contrast, \name introduces a broader range of code retrieval tasks and a unified evaluation framework for standardized benchmarking.

\noindent
\textbf{Retrieval Models.}  Dense retrievers, which encode text into high-dimensional representations, are central to information retrieval (IR). Key methods include DPR~\citep{karpukhin-etal-2020-dense}, Contriever~\citep{izacard-etal-2022-contriever}, E5~\citep{wang2022-e5}, GTE~\citep{li2023-gte}, BGE~\citep{xiao2023-bge}, and BGE-M3~\citep{bai2024m3}, with recent advances leveraging LLMs, such as E5-Mistral~\citep{wang2024improving} and OpenAI-Ada-002
% \footnote{\url{https://platform.openai.com/docs/guides/embeddings}
% }
.

However, most research focuses on QA datasets like MS-Marco~\citep{bajaj2018ms}, with limited work on code retrieval. While models like CodeBERT~\citep{feng-etal-2020-codebert}, Voyage-Code-002, and UniXCoder~\citep{guo-etal-2022-unixcoder} exploit programming language structures, neural retrieval for code remains underexplored. By providing benchmarks for code retrieval , we aim to stimulate innovation and advance state-of-the-art techniques in this area.
\section{The \name Benchmark}
%\vspace{-0.8em}

\setlength{\tabcolsep}{1.5pt}
\begin{table*}[t!]
    \renewcommand{\arraystretch}{0.8}
    \small
    \centering
    \caption{\textbf{Statistics of datasets} in \name benchmark. \# is the quantity of query/corpus instances. $L_{(\cdot)}$ refers to the average numbers of words per query/corpus. Datasets marked by $^\dag$ are created by us.}
    
    \resizebox{\textwidth}{!}{
    
    \begin{tabular}{ c  c  l  l  m{1.7cm}  c  c  c  c }
        \toprule
           \textbf{Main Task} &\textbf{Sub Task} & \textbf{Domain} & \textbf{Dataset} &  \textbf{Language} & \begin{tabular}[c]{@{}c@{}} \textbf{\#Query} \\ (train/dev/test) \end{tabular} & \textbf{\#Corpus} & \textbf{$L_{\text{Query}}$} & \textbf{$L_{\text{Corpus}}$} \\
         \midrule

    \multirow{6}{*}{ \begin{tabular}[c]{@{}c@{}} Text-to-Code \\ Retrieval \end{tabular}} 
    & \begin{tabular}[c]{@{}c@{}} Code Contest \\ Retrieval \end{tabular} 
    & \begin{tabular}[c]{@{}c@{}} Code \\ Contest \end{tabular} & APPS & py &  5k/-/3.8K & 9K & 1.4K & 575\\ \cmidrule(lr){2-9}
    & \begin{tabular}[c]{@{}c@{}} Web Query to \\Code Retrieval \end{tabular} 
    & \begin{tabular}[c]{@{}c@{}}  Web \\ query \end{tabular} & CosQA & py & 19k/-/500 & 21K & 37 & 276\\ \cmidrule(lr){2-9}
    & \begin{tabular}[c]{@{}c@{}} Text to SQL \\ Retrieval \end{tabular} 
    & Database & \begin{tabular}[c]{@{}l@{}} Synthetic \\ Text2SQL \end{tabular} & sql & 100k/-/6K & 106K & 83 &127 \\ \midrule

    \begin{tabular}[c]{@{}c@{}} Code-to-Text \\ Retrieval \end{tabular}
    & \begin{tabular}[c]{@{}c@{}} Code Summary \\ Retrieval \end{tabular} &\begin{tabular}[c]{@{}l@{}} Github \\ Fuctions \end{tabular} & CodeSearchNet & \begin{tabular}[c]{@{}l@{}}go, java, js\\ php, py, ruby \end{tabular} & 905k/41k/53K &1M &594 &156 \\ \midrule

    \multirow{6}{*}{ \begin{tabular}[c]{@{}c@{}} Code-to-Code \\ Retrieval \end{tabular}} & \begin{tabular}[c]{@{}c@{}}   Code Context \\ Retrieval \end{tabular} & \begin{tabular}[c]{@{}l@{}} Github \\ Fuctions \end{tabular} & \begin{tabular}[c]{@{}l@{}}  CodeSearchNet \\ -CCR$^\dag$ \end{tabular} & \begin{tabular}[c]{@{}l@{}}go, java, js \\ php, py, ruby\end{tabular} & 905k/41k/53K & 1M & 154 & 113 \\ \cmidrule(lr){2-9}
    & \multirow{3}{*}{ \begin{tabular}[c]{@{}c@{}} Similar Code \\ Retrieval \end{tabular}}  & \begin{tabular}[c]{@{}l@{}} Deep \\ Learning \end{tabular} & \begin{tabular}[c]{@{}l@{}} CodeTrans \\ Ocean-DL \end{tabular} & py & 564/72/180 & 816 & 1.6K & 1.5K \\ \cmidrule(lr){3-9}
    &  & Contest & \begin{tabular}[c]{@{}l@{}} CodeTrans \\ Ocean-Contest \end{tabular} &  c++, py & 561/226/446 & 1K &770 & 1.5K \\ \midrule

    \multirow{7}{*}{\begin{tabular}[c]{@{}c@{}} Hybrid Code \\ Retrieval \end{tabular} } 
    & \multirow{4}{*}{ \begin{tabular}[c]{@{}c@{}} Single-turn \\ Code QA \end{tabular} } & \begin{tabular}[c]{@{}l@{}} Stack \\ Overflow \end{tabular} & \begin{tabular}[c]{@{}l@{}} StackOverflow \\ QA$^\dag$ \end{tabular} & miscellaneous & 13k/3k/2K & 20K & 1.4K & 1.2K \\
    \cmidrule(lr){3-9}
    &  & \begin{tabular}[c]{@{}l@{}} Code \\ Instruction \end{tabular} & \begin{tabular}[c]{@{}l@{}} CodeFeedBack \\ -ST \end{tabular} & \begin{tabular}[c]{@{}l@{}} html, c, css, sql\\ js, sql, py, shell\\ ruby, rust, swift\end{tabular} & 125k/-/31K &156K & 722 &1.5K \\ \cmidrule(lr){2-9}
    & \begin{tabular}[c]{@{}c@{}} Multi-turn \\ Code QA \end{tabular} & \begin{tabular}[c]{@{}l@{}} Code \\ Instruction \end{tabular} & \begin{tabular}[c]{@{}l@{}} CodeFeeback \\ -MT \end{tabular} & miscellaneous & 53k/-/13K & 66K & 4.4K & 1.5K \\
    
    \bottomrule
    
    \end{tabular}}
    \label{tab:dataset_stats}
\end{table*}

\subsection{Desiderata}
\vspace{-0.4em}
\name provides a one-stop zero-shot evaluation benchmark for diverse code retrieval tasks, ensuring comprehensive assessment through well-curated tasks and datasets. To streamline evaluation, \name includes a one-click, end-to-end pipeline with three key features:  

(1) \textbf{Diversity}: \name assesses models across 4 primary tasks, 8 sub-tasks, 10 datasets, and 14 programming languages (Figure~\ref{fig:coir_overview}).  (2) \textbf{Usability}: Unlike traditional evaluations that require manual coding and result collection, \name offers an automated pipeline for both open-source and proprietary models, supporting metrics such as nDCG, precision, recall, and MAP. Results are stored in \texttt{JSON} format for easy access.  (3) \textbf{Overfitting Mitigation}: Many models overfit to benchmarks like CodeSearchNet, leading to inflated performance with limited generalization. \name alleviates this by incorporating diverse tasks and datasets, offering a more reliable evaluation.  Dataset statistics are in Table~\ref{tab:dataset_stats}, with preparation details in \ref{appendix:dataset_details}.

% \vspace{-0.4em}
\subsection{Overview of \name Tasks}
In this section, we present an overview of each task in \name. For each task-specific dataset, we manually inspect and filter out instances that lack valid answers, exhibit ambiguity, contain irrelevant information, etc. The rationale for selecting each dataset, along with the detailed manual filtering process, is provided in Appendix~\ref{appendix:dataset_details}.

\subsubsection{Main Task I: Text-to-Code Retrieval}
\textbf{Code Contest Retrieval.} 
Code contest retrieval involves retrieving relevant code solutions for coding problems described in natural language, a challenging task due to the complexity of language and code, as well as the gap between human and machine languages. For this, we use the APPS dataset~\citep{Hendrycks-etal-apps}, a diverse collection of problems from platforms like Codewars, AtCoder, Kattis, and Codeforces.

\textbf{Web Query Code Retrieval.} 
Web query code retrieval focuses on retrieving relevant code snippets based on concise web queries, typically just a few words. For this task, we use the CosQA~\citep{huang-etal-2021-cosqa} dataset, containing 20.6k labeled pairs of textual queries and Python functions.

\textbf{Text-to-SQL Retrieval.} 
Text-to-SQL is a key task in code generation, requiring models to generate SQL queries from natural language questions. We use the Synthetic Text-to-SQL dataset~\citep{gretel-synthetic-text-to-sql-2024}, the largest and most diverse synthetic dataset, with around 106k examples.

\vspace{-0.4em}
\subsubsection{Main Task II: Code-to-Text Retrieval}
\vspace{-0.4em}
\textbf{Code Summary Retrieval.} 
%Code Summary Retrieval measures how effective the model is in retrieving related code summaries or descriptions when given a code snippet. 
Code summary retrieval evaluates a model's ability to use code to retrieve code summaries or annotations. For this task, we employ the CodeSearchNet dataset~\citep{husain2019codesearchnet}, which consists of numerous code functions accompanied by code comments. This dataset spans six programming languages and includes over one million documents, providing a rich source of information for evaluating model performance.

\vspace{-0.4em}
\subsubsection{Main Task III: Code-to-Code Retrieval}
\vspace{-0.4em}

\textbf{Code Context Retrieval.} 
Code context retrieval is the task of retrieving the most relevant code segment that completes a given initial segment of code. This task is critical for code completion purposes. Here we modify the original CodeSearchNet dataset~\citep{husain2019codesearchnet} to better suit our needs. Specifically, for each code snippet or function in CodeSearchNet, we randomly divide the code into two segments: (1) the initial segment serves as our query, and (2) the remaining segment forms the target corpus to be retrieved for this query. The length of each query is uniformly and randomly selected to comprise between $40\%$ and $70\%$ of the total number of characters in the original code.

\textbf{Similar Code Retrieval.}
In the task of similar code retrieval, the primary objective is to assess a model's ability to retrieve similar code snippets. Specifically, given a code snippet in one programming language or deep learning framework, the model needs to retrieve semantically equivalent code in a different language or framework. We utilize the CodeTransOcean dataset~\citep{yan-etal-2023-codetransocean} for this purpose, creating two sub-datasets for similar code retrieval. These sub-datasets are named ``CodeTransOcean-DL'' and ``CodeTransOcean-Contest'', with the suffix indicating their respective domains. The ``CodeTransOcean-DL'' subset contains code written in different deep learning frameworks within the same programming language, such as TensorFlow and Paddlepaddle. Whereas the ``CodeTransOcean-Contest'' subset includes code written in different programming languages, such as Python and C++, for the same coding contest or competitive programming problem.

\vspace{-0.4em}
\subsubsection{Main Task IV: Hybrid Code Retrieval}
\vspace{-0.4em}

\textbf{Single-turn Code Question Answer Retrieval.}
In single-turn code question-answering (QA), a retrieval model is required to find the corresponding answer for a given natural language question. 
Both the question and the answer typically consist of a mix of text and code snippets. For this task, we use two code QA datasets: StackOverflow QA~\footnote{\url{https://www.kaggle.com/datasets/stackoverflow/stacksample/data}} and CodeFeedQA~\citep{zheng2024opencodeinterpreter}. The StackOverflow QA is derived from the original StackOverflow dataset by pairing questions with their highest upvoted answers, resulting in $19,931$ pairs. Additionally, we sampled $1,202$ query instances to validate the retrieval model's performance. The CodeFeedQA is a synthesized code instruction dataset generated by LLMs. From this dataset, we sampled $20\%$ of the queries to assess the model's performance.

\textbf{Multi-turn Code Question Answer.} 
In multi-turn code question-answer retrieval, the retrieval model must effectively utilize the context from multiple dialogue turns to accurately retrieve the answer for the subsequent turn. This task is challenging due to the extensive dialogue context, which can exceed $4,000$ tokens, whereas most current retrieval models are limited to a context length of $512$ tokens. Specifically, we employ the CodeFeedback multi-turn question-answer dataset, with dialogues generated by LLMs. Our test set comprises a total of $13,227$ queries and $66,383$ corpus.

\begin{figure*}[tb]
    \begin{minipage}{.4\linewidth}
        \includegraphics[width=1\textwidth]{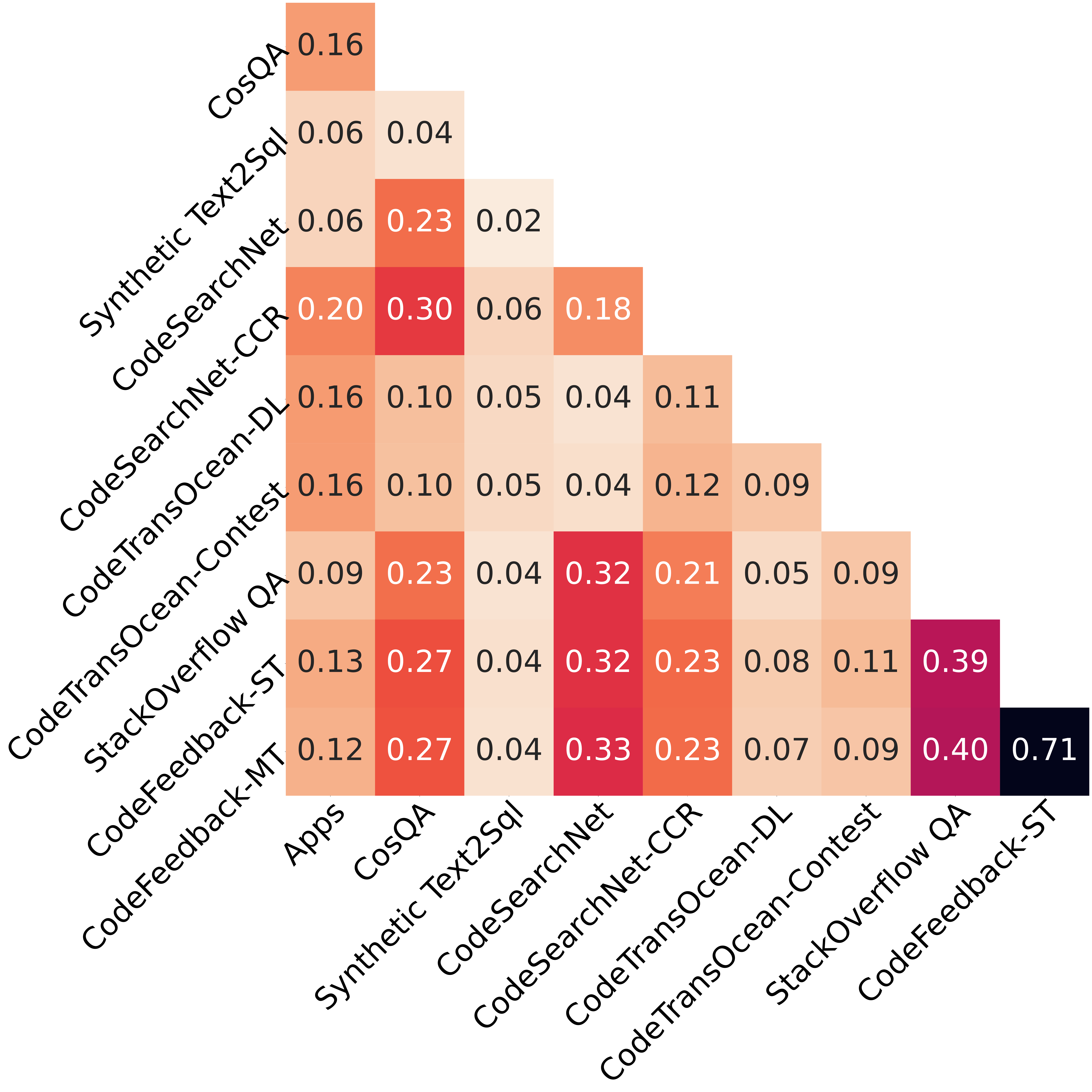}
        % \label{fig:coir_heatmap}
    \end{minipage} 
    \hspace{0.1cm}
    \begin{minipage}{.50\linewidth}
        \includegraphics[width=1\textwidth]{figures/distribution.pdf}
    \end{minipage}%
    \vspace{-1em}
    \caption{Jaccard similarity and Programming Language Distributions of \name.}
    \label{fig:coir_distribution}
\end{figure*}

\vspace{-0.4em}
\subsection{Dataset and Diversity Analysis}
\vspace{-0.4em}
In addition to the multitude of tasks, the datasets in \name also encompass a broad array of programming languages, such as Python, Java, and SQL, each featuring unique attributes.
The distribution of these programming languages is long-tailed, as depicted in the bar graph in Figure~\ref{fig:coir_distribution}.
Despite this, the datasets maintain diversity, originating from varied sources including code contest websites, GitHub repositories, StackOverflow responses, etc. 
To quantify the diversity of \name, we calculated the weighted Jaccard similarity scores on unigram word overlap across all dataset pairs, displayed in a heatmap in Figure~\ref{fig:coir_distribution} (see Appendix~\ref{sect:weighted_jaccard_similarity} for calculation details). 
The heatmap reveals generally low Jaccard similarity scores among dataset pairs, with notable exceptions being CodeFeedback Single-Turn (CodeFeedback-ST) and CodeFeedback Multi-Turn (CodeFeedback-MT), which are derived from the same domain. 
This signifies the challenge of \name as a benchmark. 
For optimal performance, a method must not only excel in major programming languages but also exhibit robust generalization ability across various domains.

\vspace{-0.4em}
\subsection{\name Evaluation Software }
\vspace{-0.4em}

To advance the evaluation of code retrieval capabilities, we introduce a streamlined, user-friendly Python framework for the \name benchmark evaluation. This framework is installable via pip using the command \texttt{pip install} and features a straightforward script that evaluates model performance across multiple datasets, outputting results in \texttt{JSON} format. Meanwhile, \name is compatible with several popular open-source frameworks, including HuggingFace and Sentence-Transformers, as well as API-based models such as OpenAI-Ada-002 and Voyage-Code-002. In line with BEIR and MTEB, all datasets have been standardized into a uniform format, facilitating the use of MTEB and BEIR frameworks for evaluating \name.

% \vspace{-0.8em}
\section{Experiment Setup}
% \vspace{-0.8em}
In this section, we evaluate and analyze the performance of the current state-of-the-art retrieval models on the eight subtasks of the \name benchmark.
More details can be found in Appendix~\ref{appendix:imple_details}.
% \subsection{Experimental Setup}

\textbf{Benchmarked Models.}    To comprehensively evaluate the capabilities of various state-of-the-art retrieval models for code retrieval tasks, we select 10 different retrieval models. 
For sparse retrieval, we utilized BM25~\citep{bm25}. 
For open-source dense models, we included E5-Base~\citep{wang2022-e5}, GTE-Base~\citep{li2023-gte}, BGE-Base~\citep{xiao2023-bge}, Contriever~\citep{izacard-etal-2022-contriever}, E5-Mistral~\citep{wang2024improving}, BGE-M3~\citep{chen2024bge}, and UniXcoder~\citep{guo-etal-2022-unixcoder}. For proprietary models, we included OpenAI-Ada-002~\footnote{\url{https://platform.openai.com/docs/guides/embeddings}} and Voyage-Code-002~\footnote{\url{https://blog.voyageai.com/2024/01/23/voyage-code-2-elevate-your-code-retrieval/}}. Notably, both UniXcoder and Voyage-Code-002 have been pre-trained on code data, making them robust baselines for comparison.

\textbf{Implementation Details.} For all open-source dense retrieval models, we set the input length of both queries and the corpus to 512. For proprietary models such as Voyage-Code-002, due to the constraint of tokens per minute (TPM), we set the length of the queries to 256 and the corpus to 512. Following previous work~\cite{wang2024improving}, we utilize (i) last token pooling for E5-Mistral, and (ii) mean pooling for other models, to obtain semantic vector representations. Cosine similarity is employed to calculate the similarity between the query and corpus for ranking.  Following BEIR~\citep{thakur2021beir}, we use NDCG@10 to evaluate the performance of retrieval models. For comprehensive evaluation, our scripts also provide metrics such as MAP, Recall, and Precision.

% \textbf{Evaluation Metric.}
% In code retrieval tasks, selecting an appropriate evaluation metric is essential for both assessing ranking sensitivity and relevance accuracy. Building on prior research~\citep{pmlr-v30-Wang13}, Normalized Discounted Cumulative Gain (NDCG) emerges as a superior metric. It effectively addresses the limitations of rank-insensitive measures such as precision and recall, as well as binary metrics like MRR and MAP that fail to account for varying relevance levels. Specifically, NDCG considers not only the order of retrieved items but also their relevance intensity. Following BEIR~\citep{thakur2021beir}, we use NDCG@10 to evaluate the performance of retrieval models. For comprehensive evaluation, our scripts also provide metrics such as MAP, Recall, and Precision.

% \usepackage{booktabs}
\begin{table*}[h]
\centering
\caption{NDCG@10 score of various retrievers on \name. The best score is marked in boldface.}
\setlength{\tabcolsep}{3pt} 
    \resizebox{\textwidth}{!}{
    \begin{tabular}{l | ccc | c | ccc | ccc | c}
        \toprule
        \multicolumn{1}{l}{\textbf{Task ($\rightarrow$)}} &
        \multicolumn{3}{c}{Text-to-Code} &
        \multicolumn{1}{c}{Code-to-Text} &
        \multicolumn{3}{c}{Code-to-Code} &
        \multicolumn{3}{c|}{Hybrid Code} &
        \multicolumn{1}{c}{\multirow{3}{*}{Avg}}
        \\ 
        \cmidrule(lr){1-1}
        \cmidrule(lr){2-4}
        \cmidrule(lr){5-5}
        \cmidrule(lr){6-8}
        \cmidrule(lr){9-11}

        \multicolumn{1}{l|}{\multirow{2}{*}{\textbf{Model (param.) $\downarrow$}}} &
        \multicolumn{1}{c}{\multirow{2}{*}{Apps}} &
        \multicolumn{1}{c}{\multirow{2}{*}{CosQA}} &
        \multicolumn{1}{c|}{Synthetic} &
        \multicolumn{1}{c|}{Code} &
        \multicolumn{1}{c}{Code} &
        \multicolumn{2}{c|}{\underline{CodeTrans}} &
        \multicolumn{1}{c}{StackOver} &
        \multicolumn{2}{c|}{\underline{CodeFeedBack}} &
         \\
        % \cmidrule(lr){7-8}
        % \cmidrule(lr){10-11}
        & & & Text2sql & SearchNet & SN-CCR & -Contest & -DL & Flow QA & -ST & -MT &  \\
        
        \midrule
        BM25  & 0.95 & 13.96 & 16.92 & 26.75 & 34.69 & 50.13 & 8.69 & 56.80 & 54.32 & 34.73 & 29.79 \\

        Contriever (110M) & 5.14 & 14.21 & 45.46 & 34.72 & 35.74 & 44.16 & 24.21 & 66.05 & 55.11 & 39.23 & 36.40 \\
        E5-base (110M) & 11.52 & 32.59 & 52.31 & 67.99 & 56.87 & 62.50 & 21.87 & 86.86 & 74.52 & \textbf{41.99} & 50.90 \\
        BGE-Base (110M) & 4.05 & \textbf{32.76} & 45.59 & 69.60 & 45.56 & 38.50 & 21.71 & 73.55 & 64.99 & 31.42 & 42.77 \\
        GTE-Base (110M) & 3.24 & 30.24 & 46.19 & 43.35 & 35.50 & 33.81 & 28.80 & 62.71 & 55.19 & 28.48 & 36.75 \\
        UniXcoder (123M) & 1.36 & 25.14 & 50.45 & 60.20 & 58.36 & 41.82 & 31.03 & 44.67 & 36.02 & 24.21 & 37.33 \\
        BGE-M3 (567M) & 7.37 & 22.73 & 48.76 & 43.23 & 47.55 & 47.86 & 31.16 & 61.04 & 49.94 & 33.46 & 39.31 \\
        E5-Mistral (7B) & 21.33 & 31.27 & 65.98 & 54.25 & 65.27 & \textbf{82.55} & \textbf{33.24} & \textbf{91.54} & \textbf{72.71} & 33.65 & 55.18 \\
        OpenAI-Ada-002 & 8.70 & 28.88 & 58.32 & 74.21 & 69.13 & 53.34 & 26.04 & 72.40 & 47.12 & 17.74 & 45.59 \\
        Voyage-Code-002 & \textbf{26.52} & 29.79 & \textbf{69.26} & \textbf{81.79}  & \textbf{73.45} & 72.77 & 27.28 & 87.68 & 65.35 & 28.74 & \textbf{56.26} \\
        
        \bottomrule
    \end{tabular}}
    \label{tab:results}
    \vspace{-0.6em}
\end{table*}

% \vspace{-0.8em}
\section{Results and Analysis}

\subsection{Main Results}
% \vspace{-0.5em}

Table~\ref{tab:results} showcases the performance of existing embedding models on the \name benchmark. From the table, several  conclusions can be drawn:

\textbf{No Single Model Dominates Across All Tasks.}
Despite Voyage-Code-002 achieving the highest mean score of 56.26 on the benchmark, it does not universally surpass other models in every task. Models like E5-base and E5-Mistral exhibit superior performance in specific retrieval sub-tasks. This highlights the \name benchmark's complexity and suggests that different models excel in different facets, catering to varied task requirements.

\textbf{Demonstration of Benchmark Diversity and Challenge.}
Models achieving top performance in mainstream text retrieval benchmarks, such as E5-Mistral, do not consistently excel across \name sub-tasks. For example, E5-Mistral scores highly in CodeTrans-Contest and StackOverflow QA but shows only median performance in APPS and CosQA for Text-to-Code tasks, and falls behind E5-base in CodeFeedBack-MT. This variation underscores the \name benchmark's challenge and the need for adaptive, robust code retrieval capabilities.

\textbf{Effectiveness of Code-Specific Optimization.}
Models specialized for code retrieval, like Voyage-Code-002, achieve strong results on \name, highlighting the impact of tailored optimization. Voyage-Code-002 excels in both Text-to-Code and Code-to-Text tasks, demonstrating the benefits of domain-specific training.  

\begin{figure}[htbp] 
    \centering
    \includegraphics[width=0.9\textwidth, trim={0.0cm 0.15cm 0.0cm 0.38cm}, clip]{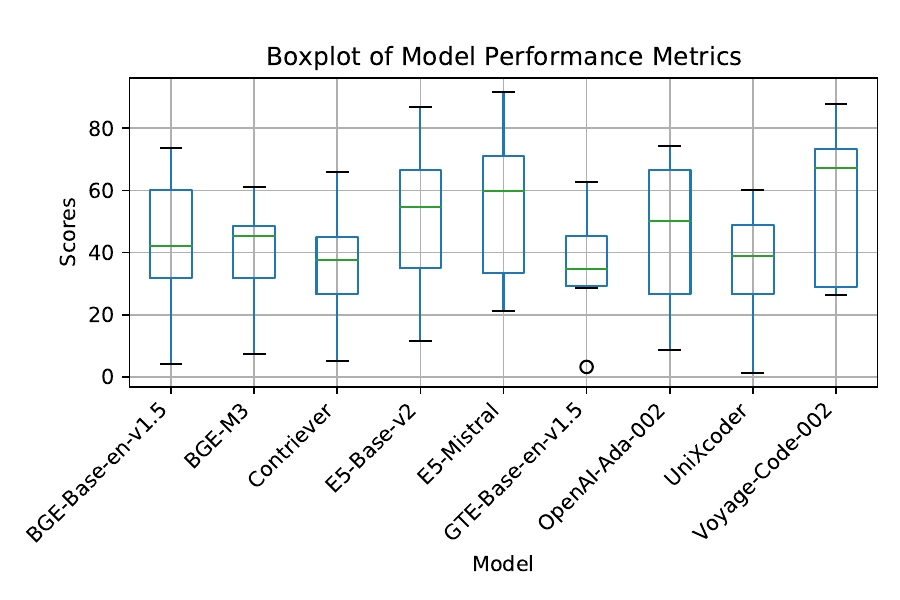} 
    % \vspace{-0.8em}
    \caption{The box plots of all model performances on  \name benchmark.} 
    \label{fig:blox} 
    \vspace{-1.5em}
\end{figure}

Figure~\ref{fig:blox} presents box plots for each model. Voyage-Code-002 achieves the highest average performance but with high variance, suggesting weaker generalization. In contrast, BGE-M3 shows better robustness with the lowest variance despite its moderate performance. This highlights the CoIR dataset's challenge and its value as a benchmark for code retrieval research.
% Models tailored specifically for code retrieval, like Voyage-Code-002, achieve impressive results on \name. This indicates that specialized optimization can significantly boost performance. Voyage-Code-002 excels in Text-to-Code and Code-to-Text tasks, underscoring the advantages of specific training and optimization for code retrieval.

% We also report the box plots for each model, as shown in Figure~\ref{fig:blox}. We can observe that although Voyage-Code-002 exhibits the highest average performance, it also has the largest variance. This suggests that Voyage-Code-002 struggles with generalization. In contrast, while BGE-M3 demonstrates only average performance, it has the smallest variance, indicating that BGE-M3 may possess better generalization capabilities. This observation highlights the challenging nature of the CoIR dataset, as no single model dominates across all datasets. Consequently, CoIR serves as a valuable benchmark for driving progress in the field of code retrieval.

\subsection{Analysis of Retrieval Efficiency}
In practical industrial settings, efficiency is also a critical factor for evaluating retrieval model. Factors such as embedding latency for query/corpus, retrieval latency from all embeddings, and the physical memory footprint of the index are critical. In this section, we analyze the efficiency of the current retrieval model using a portion of the \name dataset, designated as CodeFeedBack-ST, which comprises 156k corpus and 31k queries.
For \textbf{embedding latency}, we record the time taken by each model to process a batch of data. This time was then divided by the number of samples in the batch to derive the average embedding time per sample. 
For \textbf{retrieval latency}, we measure the averaged retrieval time of all queries, where each query is tasked with retrieving 1,000 documents.
The latency results are presented in Table~\ref{tab:effiency}.

\setlength{\tabcolsep}{3pt}
\renewcommand{\arraystretch}{0.9}
\begin{table}[htbp]{}{} % 调整宽度
  \resizebox{1\textwidth}{!}{ % 调整表格大小
\begin{tabular}{@{}m{0.2cm}cc|c|c|c@{}}
\toprule
\multicolumn{3}{c|}{\begin{tabular}[c]{@{}c@{}} CodeFeedBack-ST \\ 156k corpus \& 31k queries \end{tabular}}          & \begin{tabular}[c]{@{}c@{}}Embed \\ Latency \end{tabular} & \begin{tabular}[c]{@{}c@{}}Retrieval \\ Latency \end{tabular} & Index  \\ \midrule
Rank & \multicolumn{1}{c|}{Model}           & Dim. & GPU   & GPU   & Size   \\ \midrule
9    & \multicolumn{1}{c|}{Contriever}      & 768  & 7.8ms & 38.1µs  & 0.3G  \\
3    & \multicolumn{1}{c|}{E5-Base}         & 768  & 7.4ms & 38.1µs  & 0.3G  \\
5    & \multicolumn{1}{c|}{BGE-Base}        & 768  & 7.6ms & 38.1µs  & 0.3G  \\
8    & \multicolumn{1}{c|}{GTE-Base}        & 768  & 7.8ms & 38.1µs  & 0.3G  \\
7    & \multicolumn{1}{c|}{UniXcoder}     & 768  & 7.7ms   & 38.1µs  & 0.3G  \\
6    & \multicolumn{1}{c|}{BGE-M3}          & 1024 & 31.4ms & 42.9µs & 0.6G  \\
2    & \multicolumn{1}{c|}{E5-Mistral}   & 4096 & 1840ms & 115.5µs  & 2.3G \\
4    & \multicolumn{1}{c|}{OpenAI-Ada-002}  & 1536 & - & 56.8µs  & 0.9G  \\
1    & \multicolumn{1}{c|}{Voyage-Code-002} & 1536 & - & 56.8µs  & 0.9G  \\ 
\bottomrule
\end{tabular}
  }
  % \vspace{-0.8em}
  \caption{The average embedding/retrieval latency per instance using CodeFeedBack-ST. We retain models that have rankings in both BEIR and \name. }
  
  \label{tab:effiency}
  \vspace{-0.8em}
\end{table}

\textbf{Trade-off between Accuracy and Latency.}
High-performing models like E5-Mistral with excellent accuracy, suffer from noticeable embedding latency. Specifically, E5-Mistral's average embedding latency is 1840ms per sample, significantly higher than other evaluated models. This discrepancy highlights the common trade-off in retrieval systems: higher accuracy often comes at the expense of increased latency.

\textbf{Index Size Considerations.} 
Index sizes vary significantly across different models. Simpler models such as Contriever, E5-Base, BGE-Base, GTE-Base, and UniXcoder have relatively small index sizes, around 0.3GB. In contrast, more complex models like E5-Mistral, OpenAI-Ada-002, and Voyage-Code-002 have larger index sizes ranging from 0.6GB up to 2.3GB. This suggests a trade-off between accuracy and memory demands. Advanced models with better performance is at the cost of more memory, which could be a limitation in resource-constrained environments.

\subsection{Input Length Impact on Code Retrieval}
This section examines how input length affects the performance of code retrieval tasks. We analyze results using four datasets: 
CodeFeedBack-MT, CodeTransOcean-DL, APPS, and StackOverflow QA. 
Each dataset has an average query and corpus length that exceeds 1,000 words.
% Each of these datasets features an average context length for both queries and corpus exceeding 1,000 tokens. 
We utilize two models, GTE and BGE-M3, both of which have been optimized for long documents and support context length of 8k tokens. We experiment on two settings, with input length capped at 512 and 4,096 respectively. Note that tokens beyond the cap will be truncated. We report the results in Table~\ref{tab:coir_length}.

\renewcommand{\arraystretch}{0.9}
\begin{table}[htbp]{}{} % 调整宽度
   \resizebox{1\textwidth}{!}{ % 调整表格大小
\setlength{\tabcolsep}{2pt}
\begin{tabular}{@{}ccccc@{}}
\toprule 
% \multicolumn{1}{c}{\multirow{2}{*}{Model}} &
% \multicolumn{1}{c}{CodeFeed} &
% \multicolumn{1}{c}{CodeTrans} &
% \multicolumn{1}{c}{\multirow{2}{*}{Apps}} &
% \multicolumn{1}{c}{StackOver} \\
%  & Back-MT & Ocean-DL & & FlowQA \\
 \multicolumn{1}{c}{Model} &
\multicolumn{1}{c}{Code} &
\multicolumn{1}{c}{Code} &
\multicolumn{1}{c}{\multirow{2}{*}{APPS}} &
\multicolumn{1}{c}{Stack} \\
(input \#tokens) & FB-MT & TO-DL & & OF-QA \\
\midrule
GTE (512)    & 28.48          & 28.80            & 3.24 & 62.71          \\
GTE (4k)     & 51.32          & 27.33            & 5.08 & 78.63          \\
BGE-M3 (512)  & 33.46          & 31.16           & 7.37 & 61.04         \\
BGE-M3 (4k)  & 27.49          & 32.75            & 6.80 & 56.53          \\ \bottomrule
\end{tabular}
}
  \vspace{-0.7em}
  \caption{Effects of different input length.
  \vspace{-0.7em}
  \label{tab:coir_length}}
\end{table}
  % \vspace{-1.2em}
\textbf{Impact of Input Length on Model Performance.} For GTE model, extending the input length from 512 to 4,096 shows notable improvements in retrieval performance across most datasets. Specifically, retrieval scores in CodeFeedBack-MT and StackOverflow QA increase significantly from 38.20 to 51.32, and from 64.36 to 78.63, respectively. In contrast, the BGE-M3 model shows inconsistent results. 
% While the performance for some datasets such as CodeTransOcean-DL increases with the extended input length (from 31.16 to 32.75), the performance for others, like CodeFeedBack-MT, decreases (from 33.46 to 27.49).
CodeTransOcean-DL shows a slight improvement in scores from 31.16 to 32.75, whereas scores for CodeFeedBack-MT fall from 33.46 to 27.49. One possible reason can be that:
% We hypothesize that 
although BGE-M3 has been optimized for long documents, the significant differences between code data and text data result in a performance degradation as the document length increases.

% \subsection{Comparison of Rankings in \name and BEIR Benchmarks}
\vspace{-0.8em}
\subsection{Comparison of \name and BEIR Rankings}  
\begin{table}[htbp]
  \centering
    \begin{tabular}{@{}ccc@{}}
      \toprule
      \multicolumn{1}{c}{\multirow{2}{*}{Model}} &
      \multicolumn{2}{c}{Rank} \\
      \cmidrule{2-3}
      & \multicolumn{1}{c}{CoIR} & \multicolumn{1}{c}{BEIR} \\
      \midrule
      Contriever     & 7         & 7         \\
      GTE-Base       & 6         & 2         \\
      BGE-M3         & 5         & 6         \\
      BGE-Base       & 4         & 3         \\
      OpenAI-Ada-002 & 3         & 5         \\
      E5-Base        & 2         & 4         \\
      E5-Mistral     & 1         & 1         \\ 
      \bottomrule
    \end{tabular}
  \caption{Retriever rankings in CoIR  and BEIR.}
  \label{tab:rank}
  \vspace{-0.9em}
\end{table}

% Notably, UniXcoder and Voyage-Code-2, do not have corresponding rankings in BEIR.  
This section evaluates seven retrieval models across \name and BEIR, adjusting rankings for consistency while excluding voyage-code-002 and UniXcoder due to their code-specific pretraining. We analyze key ranking shifts and patterns.  

\textbf{Key Insights.}  
E5-Mistral consistently ranks first in both benchmarks, demonstrating robust performance in text and code retrieval. However, GTE-Base drops from 2$^{\text{nd}}$ in BEIR to 6$^{\text{th}}$ in \name, indicating that strong text retrieval does not ensure effective code retrieval. Conversely, E5-Base ranks higher in \name than in BEIR, suggesting better adaptation to code-related tasks.  

These results underscore the necessity of specialized benchmarks like \name. While some models maintain stable rankings, others exhibit notable shifts, highlighting distinct challenges of code retrieval that general text benchmarks may overlook.

\subsection{Overfitting in CodeSearchNet and How \name Helps}
\begin{figure}[htbp] 
    \centering
    \includegraphics[width=0.9\textwidth, trim={0.0cm 0.15cm 0.0cm 0.38cm}, clip]{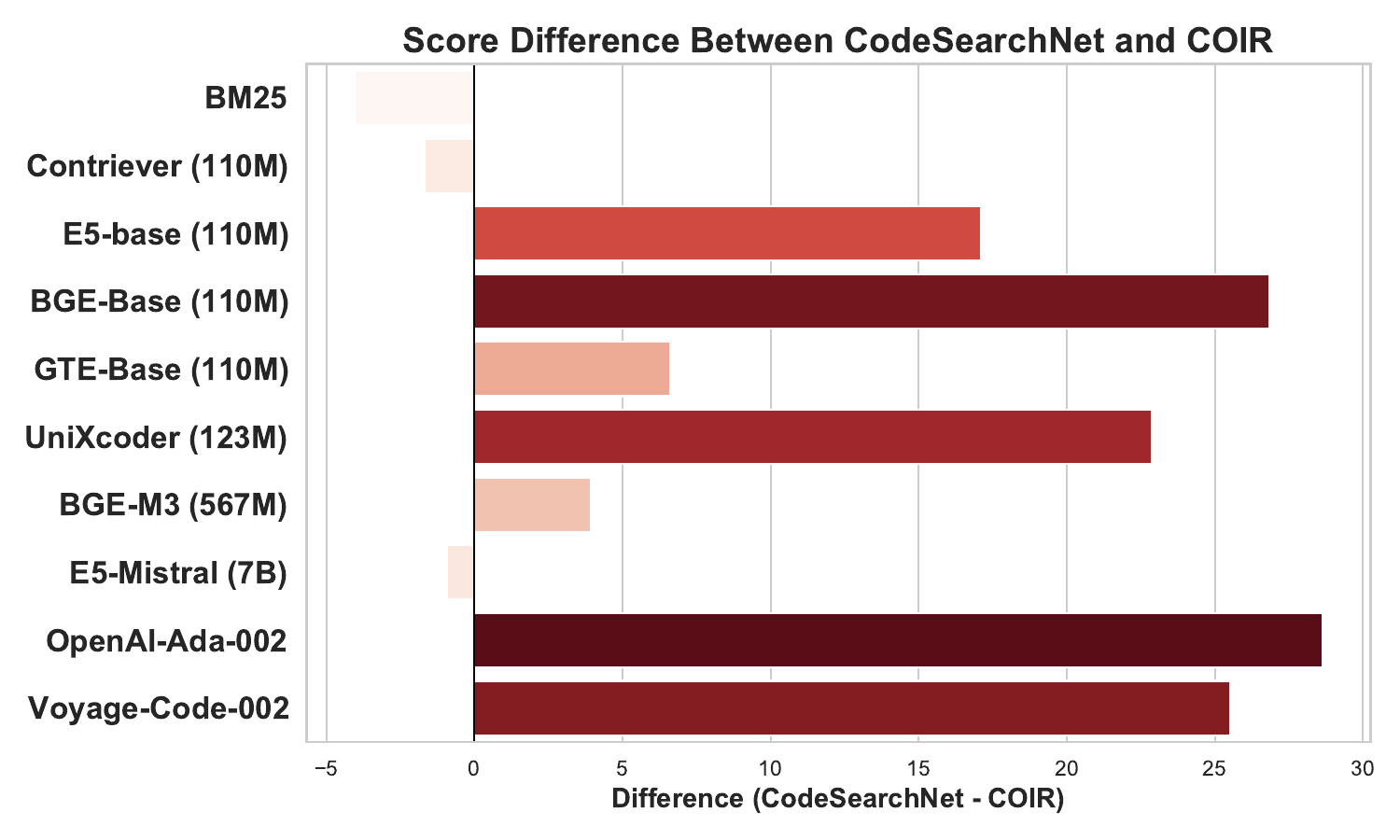} 
    % \vspace{-0.8em}
    \caption{Overfitting in CodeSearchNet: Evidence from Score Differences. The values in the bar chart represent the scores of CodeSearchNet minus the scores of \name.} 
    \label{fig:score_diff} 
    \vspace{-0.8em}
\end{figure}

Figure~\ref{fig:score_diff} compares the performance of CodeSearchNet and \name across various retrieval models. Most models score significantly higher on CodeSearchNet, indicating a strong overfitting tendency. Notably, OpenAI-Ada-002 and Voyage-Code-002 show large performance gaps, suggesting these models may have already overfitted to CodeSearchNet. 

In contrast, \name provides a more balanced evaluation, mitigating overfitting. Among all models, E5-Mistral (7B) achieves the \textbf{smallest performance gap} while maintaining strong scores on both benchmarks, demonstrating that \textbf{LLM-based retrieval models can effectively reduce overfitting}. These results highlight \name’s role in ensuring a more robust and generalizable benchmark.

\section{Conclusion}
In this paper, we introduce \name, a comprehensive benchmark for code retrieval. \name encompasses \textbf{4} distinct and \textbf{8} fine-grained retrieval tasks, supports \textbf{14} programming languages, and integrates \textbf{10} diverse datasets with over 2 million code snippets. Furthermore, we evaluate the performance of \textbf{10} retrieval models on \name, revealing that even state-of-the-art models struggle, underscoring its challenging nature. Additionally, our analysis suggests that many existing models have overfitted to current leaderboards, highlighting the need for more generalizable and robust retrieval approaches. Moreover, we emphasize the promise of LLM-based retrieval models as a potential direction for future advancements. With \name, we aim to foster progress in code retrieval, encouraging researchers to develop more effective and resilient models to benefit the community.

\section{Limitations}
\label{appendix:limit}
Even though we cover a wide range of tasks and domains in \name, no benchmark is perfect and each set of benchmark has its own limitations.
It is crucial to make the limitations explicit in order to better interpret (1) retrieval results on these benchmark datasets and (2) to curate a better benchmark in the future that complements existing benchmarks in the field.

\begin{enumerate}[leftmargin=*]
    \item \textbf{Multilingual Tasks:} Although we aim for a diverse retrieval evaluation benchmark, due to the limited availability of multilingual retrieval datasets
    for code information retrieval
    , all datasets covered in the \name benchmark are currently English. Future work could include multi- and cross-lingual tasks and models.
    \item \textbf{Multi-faceted Search:} 
    Due to the existing paradigm that heavily emphasizes on semantic-based information retrieval, our benchmark mainly focuses on queries that aim to benchmark retrieval performance solely based on textual information. However, real-world information needs is often complex and could possibly rely on various corpus meta-data, especially so in code datasets where meta-data could play a huge role. For instance, the versioning of programming language or software libraries could make a huge difference to whether a functioning code snippet is retrieved. In essence incorporating queries that are multi-faceted can more accurately reflect real-world informational needs.

    \item \textbf{$n$-ary Match:} 
    With each query corresponding to exactly one ground-truth corpus, we dismiss the 
    real-world informational needs where (1) a single query is could be relevant to multiple copora and (2) the informational needs of a single query can only be satisfied by multiple corpora simultaneously. 

    Dedicating a particular section of \name for $n$-nary / list-wise labels for each query would allow us to address how model perform on informational needs that are diverse and further enhance the diversity of tasks of the benchmark. 
\end{enumerate}

\section{Ethical Considerations}
We ensure that the distribution of each dataset complies with the corresponding licenses, all of which are listed below:
\begin{enumerate}
    \item[$\bullet$] APPS: Provided under ``MIT License'' for non-commercial research purposes.
    \item[$\bullet$] CodeTransOcean: Provided under Apache License 2.0 license.    
    \item[$\bullet$] CodeSearchNet: Provided under ``MIT License'' for non-commercial research purposes.
    \item[$\bullet$] CoSQA: Provided under ``MIT License'' for non-commercial research purposes.
    \item[$\bullet$] Synthetic Text2sql: Provided under Apache License 2.0 license.
    \item[$\bullet$] Code-Feedback: Provided under Apache License 2.0 license.
    \item[$\bullet$] CodeFeedback-Filtered-Instruction: Provided under Apache License 2.0 license.
    \item[$\bullet$] Stackoverflow QA: Provided under CC-BY-SA 3.0 license.

    % \item[$\bullet$] MSMARCO: Provided under ``MIT License'' for non-commercial research purposes.
    % \item[$\bullet$] FEVER, NQ, DBPedia, Signal-1M: All provided under CC BY-SA 3.0 license.
    % \item[$\bullet$] TREC-NEWS, Robust04, BioASQ: Data collection archives are under \textbf{Copyright}.
    % \item[$\bullet$] ArguAna, Touch\'e-2020: Provided under CC BY 4.0 license.
    % \item[$\bullet$] CQADupStack: Provided under Apache License 2.0 license.
    % \item[$\bullet$] SciFact: Provided under the CC BY-NC 2.0 license.
    % \item[$\bullet$] SCIDOCS: Provided under the GNU General Public License v3.0 license.
    % \item[$\bullet$] HotpotQA: Provided under the CC BY-SA 4.0 license.
    % \item[$\bullet$] TREC-COVID: Provided under the ``Dataset License Agreement''.
\end{enumerate}

% To facilitate evaluation, we also propose an accompanying python package for researchers to utilize. We anticipate that \name will advance the field of code retrieval, encouraging researchers to develop more robust retrieval models to benefit the community.

\clearpage

\bibliography{acl_latex}

\appendix

\section{Dataset Filtering and Cleaning Process}
\label{appendix:dataset_construct}
The dataset used for code retrieval tasks must be of high quality, diverse, and appropriately challenging to ensure the development of robust models. To this end, we implemented a multi-stage filtering and cleaning process that removes irrelevant, incomplete, or overly simplistic code snippets, ensuring the dataset is representative of real-world programming tasks.
\subsection{Difficulty}
\begin{table*}[htbp]
    \centering
    \begin{tabular}{lcccc}
        \toprule
        \textbf{Model (NDCG@10)} & \textbf{Human-eval} & \textbf{XcodeEval (sampled 13k)} & \textbf{mbpp} \\
        \midrule
        E5-Base-v2            & 92.41  & 89.59  & 95.43 \\
        BGE-Base-en-v1.5       & 93.45  & 88.47  & 92.41 \\
        GTE-Base-en-v1.5       & 91.54  & 87.54  & 94.45 \\
        UniXcoder              & 92.44  & 90.85  & 90.56 \\
        \bottomrule
    \end{tabular}
    \caption{Model performance comparison on Simple Dataset }
    \label{tab:simple_dataset}
\end{table*}

To ensure that each selected dataset presents an appropriate level of difficulty, we have curated a large collection of datasets. For datasets that are overly simplistic and fail to effectively differentiate between model capabilities, we have opted to discard them. The performance of models on these overly simple datasets is summarized in Table~\ref{tab:simple_dataset}.
\subsection{Deduplication}
The first step involved identifying and removing duplicate code snippets. Duplicate examples can distort the model's performance by introducing redundancy, which may lead to overfitting. Using hash-based techniques, we identified exact duplicates. This ensured that each code snippet in the dataset was unique and contributed distinct information.

\subsection{Alignment}
For code retrieval tasks, the alignment between the code and its associated natural language description is critical. We filtered out examples where comments or documentation were missing, incomplete, or irrelevant to the functionality of the code. This step ensured that the dataset only contained examples with high-quality, informative descriptions that could aid in retrieval tasks.

\section{Dataset  Details} 
\subsection{Reasons for Dataset Selection}

\begin{itemize}
    \item \textbf{APPS}: A challenging dataset where state-of-the-art retrieval models achieve an NDCG@10 of only 26.52, highlighting its difficulty.
    \item \textbf{CosQA}: The largest web-based query-to-code dataset, featuring meticulously crafted queries to enhance retrieval performance.
    \item \textbf{Synthetic Text-to-SQL}: The largest synthetic text-to-SQL dataset, enabling robust evaluation of SQL generation models.
    \item \textbf{CodeSearchNet}: A widely used code retrieval dataset covering six programming languages, with a corpus of 1 million documents.
    \item \textbf{CodeSearchNet-CCR}: Specifically designed for the Code Context Retrieval task, which is crucial for improving code completion systems.
    \item \textbf{CodeTransOcean-DL \& CodeTransOcean-Contest}: Well-suited for Similar Code Retrieval, encompassing deep learning implementations and competitive programming solutions across multiple languages.
    \item \textbf{StackOverflow QA}: Reflects real-world developer queries, making it ideal for evaluating models that retrieve answers based on complete questions.
    \item \textbf{CodeFeedBack-ST}: A high-quality synthetic QA dataset, beneficial for retrieval-augmented generation and large language models.
    \item \textbf{CodeFeedBack-MT}: Designed for Multi-turn Code QA, presenting a significant challenge due to its extensive dialogue context exceeding 4,000 tokens—far beyond the typical 512-token limit of most models.
\end{itemize}

\label{appendix:dataset_details}
Examples of queries and corpora present in each dataset can be viewed in tables \ref{tab:examples1} and \ref{tab:examples2}.
\begin{table*}[t!]
    \small
    \resizebox{\textwidth}{!}{
    \begin{tabular}{ l | l | p{7.5cm} | p{7cm} | l }
        \toprule
        \multicolumn{2}{l|}{\textbf{Dataset}} &
        \multicolumn{1}{p{7.5cm}|}{\textbf{Query}} &
        \multicolumn{1}{p{7cm}}{\textbf{Relevant-Corpus}} &
        \multicolumn{1}{|c}{\textbf{Granularity}} \\ 
        \midrule
        
       \parbox[t]{3mm}{\multirow{20}{*}{\rotatebox[origin=c]{90}{\textbf{Text Query to Code Corpus}}}} 
       
       & Apps & \multicolumn{1}{p{7.5cm}|}{You are playing a very popular game called Cubecraft. Initially, you have one stick and want to craft k torches. 
       \textit{<Text omitted for brevity>}
       % One torch can be crafted using one stick and one coal. Hopefully, you've met a very handsome wandering trader who has two trade offers:  exchange 1 stick for x sticks (you lose 1 stick and gain x sticks).  exchange y sticks for 1 coal (you lose y sticks and gain 1 coal). During one trade, you can use only one of these two trade offers. You can use each trade offer any number of times you want to, in any order. 
       % Your task is to find the minimum number of trades you need to craft at least k torches. 
       % The answer always exists under the given constraints. You have to answer t independent test cases. \textit{<Example Input and Output Omitted>}  
       For each test case, print the answer: the minimum number of trades you need to craft at least k torches. 
       The answer always exists under the given constraints. \textit{<Example Test Case Omitted>}} & 
       
       % \vspace{-\baselineskip}
\begin{lstlisting}[boxpos=t]
for haaghfj in range(int(input())):
    x,y,k = list(map(int,input().split()))
    print(k + (y * k + k - 1 +x-2) // (x - 1))
\end{lstlisting} & Function \\ \cmidrule{2-5}
       
       & CosQA & \multicolumn{1}{p{7.5cm}|}{python adjacency matrix from edge list} & \multicolumn{1}{p{7cm}|}{% \vspace{-\baselineskip}
\begin{lstlisting}[boxpos=t]
def get_adjacent_matrix(self):
    edges = self.edges
    num_edges = len(edges) + 1
    adj = np.zeros([num_edges, num_edges])
    for k in range(num_edges - 1):
        adj[edges[k].L, edges[k].R] = 1
        adj[edges[k].R, edges[k].L] = 1
    return adj
\end{lstlisting}
% \vspace{-10cm}
} & Function \\  \cmidrule{2-5}
       
       &  \multirow{2}{*}{\begin{tabular}[c]{@{}c@{}} Synthetic \\ Text2Sql \end{tabular}}
       % Synthetic Text2Sql
       & \multicolumn{1}{p{7.5cm}|}{Which buildings in the UK have both a green roof and solar panels installed?} & % \vspace{-\baselineskip}
\begin{lstlisting}[boxpos=t]
SELECT b.name 
    FROM Building b JOIN GreenRoof gr ON b.id = gr.building_id JOIN SolarPanel sp ON 
    b.id = sp.building_id 
    WHERE b.country = `UK';
\end{lstlisting} & 
       % \multirow{2}{*}{\begin{tabular}[c]{@{}c@{}} Entire \\ Code \end{tabular} }
       Entire Code
       \\  \midrule

        \parbox[t]{3mm}{\multirow{43}{*}{\rotatebox[origin=c]{90}{\textbf{Code Query to Code Corpus}}}} 
        
       & \multirow{3}{*}{\begin{tabular}[c]{@{}c@{}} CodeSe- \\ archNet \\-CCR \end{tabular}} 
        % \begin{tabular}[c]{@{}c@{}} CodeSe- \\ archNet \\-CCR \end{tabular} 
        % CodeSearchNet-CCR 
        & % \vspace{-\baselineskip}
\begin{lstlisting}[boxpos=t]
def get_cumulative_spend(key):
    """
    Get the sum of spending for this category 
    up to and including the given month.
    """
    query = (`ROUND(SUM(total_ex_vat), 2) 
        AS total ' `FROM {table} '
        `WHERE date <= "{year}-{month:02}-01" '
        `AND lot="{lot}" '
        `AND customer_sector="{sector}" '
        `AND supplier_type="{sme_large}"'.
            format(
            table=_RAW_SALES_TABLE,
            year=key.year,
            month=key.month,
\end{lstlisting} & % \vspace{-\baselineskip}
\begin{lstlisting}[boxpos=t]


    lot=key.lot,
    sector=key.sector,
    sme_large=key.sme_large))
logging.debug(query)
result = scraperwiki.sqlite.select(query)
logging.debug(result)
value = result[0][`total']
return float(result[0][`total']) if value 
is not None else 0.0
\end{lstlisting} & 
        \multirow{2}{*}{\begin{tabular}[c]{@{}c@{}} Code \\ Snippet \end{tabular}} \\
        % Code Snippet \\
        \cmidrule{2-5}
        
       & \multirow{2}{*}{\begin{tabular}[c]{@{}c@{}} CodeTrans \\ Ocean-DL \end{tabular} }
        % CodeTransOcean-DL
        & % \vspace{-\baselineskip}
\begin{lstlisting}[boxpos=t]
import tensorflow as tf
from d2l import tensorflow as d2l
net = tf.keras.models.Sequential([tf.
    keras. layers. Flatten(), tf.keras.
    layers. Dense(256, activation='relu'), tf.keras.layers.Dense(10)])
batch_s, lr, num_epochs = 256, 0.1, 10
loss = tf.keras.losses.
    SparseCategoricalCrossentropy()
trainer = tf.keras.optimizers.SGD(lr)
train_iter, test_iter = d2l.load_data_fashion_mnist(batch_s)
d2l.train_ch3(net, train_iter, test_iter, 
loss, num_epochs, trainer   
\end{lstlisting} & % \vspace{-\baselineskip}
\begin{lstlisting}[boxpos=t]
from d2l import paddle as d2l
import paddle
from paddle import nn
net = nn.Sequential(nn.Flatten(),
        nn.Linear(784, 256), nn.ReLU(),
        nn.Linear(256, 10))
for layer in net:
    if type(layer) == nn.Linear:
        weight_attr = paddle.framework.ParamAttr(
        initializer=paddle.nn.initializer.
        Normal(mean=0.0, std=0.01))
        layer.weight_attr = weight_attr
batch_size, lr, num_epochs = 256, 0.1, 10
... Code truncated for brevity
\end{lstlisting}
% \centering\textit{<Imports, Code truncated for brevity>}
 & 
        % \multirow{2}{*}{\begin{tabular}[c]{@{}c@{}} Entire \\ Code \end{tabular} }\\
        Entire Code \\ 
        \cmidrule{2-5}

        & \multirow{3}{*}{\begin{tabular}[c]{@{}c@{}} CodeTrans \\ Ocean- \\Contest \end{tabular} }
        % CodeTransOcean-Contest 
        & % \vspace{-\baselineskip}
\begin{lstlisting}[boxpos=t]
def setup():
   println(distance("kitten", "sitting"))
   
def distance(a, b):
   costs = []
   for j in range(len(b) + 1):
      costs.append(j)
   for i in range(1, len(a) + 1):
      costs[0], nw = i, i - 1
      for j in range(1, len(b) + 1):
         cj=min(1+min(costs[j],costs[j-1]),
            nw if a[i-1]==b[j-1] else nw+1)
         nw, costs[j] = costs[j], cj
   return costs[len(b)]
\end{lstlisting} & % \vspace{-\baselineskip}
\begin{lstlisting}[boxpos=t]
#include <algorithm>
#include ...
template <typename StringType>
size_t levenshtein_distance(const StringType& s1, const StringType& s2) {
    const size_t m = s1.size();
    const size_t n = s2.size();
    if (m == 0) return n;
    if (n == 0) return m;
    std::vector<size_t> costs(n + 1);
    std::iota(costs.begin(),costs.end(),0);
    size_t i =...
... Code truncated for brevity}
\end{lstlisting}
% \centering\textit{<Imports, Code truncated for brevity>}
 &
        Entire Code \\ 
        % \multirow{2}{*}{\begin{tabular}[c]{@{}c@{}} Entire \\ Code \end{tabular}} \\ 

    \bottomrule
    \end{tabular}}
    \caption{Examples of queries and relevant corpora for 6 datasets related to Text-to-Code and Code-to-Code retrieval (i.e. \emph{Apps}, \emph{CosQA}, \emph{Synthetic Text2sql}, \emph{CodeSearchNet-CCR}, \emph{CodeTransOcean-DL}, and \emph{CodeTransOcean-Contest})  in \name. For brevity, we omit or truncate some portion of the query and corpus.}
    \label{tab:examples1}
\end{table*}
\begin{table*}[t!]
    \small
    \resizebox{\textwidth}{!}{
    \begin{tabular}{l | l | p{7.5cm} | p{7cm} | l }
        \toprule
        \multicolumn{2}{l|}{\textbf{Dataset}} &
        \multicolumn{1}{p{7.5cm}|}{\textbf{Query}} &
        \multicolumn{1}{p{7cm}}{\textbf{Relevant-Corpus}} &
        \multicolumn{1}{|c}{\textbf{Granularity}} \\ 
        \midrule

        \parbox[t]{3mm}{\multirow{13}{*}{\rotatebox[origin=c]{90}{\textbf{Code Query to Code Corpus}}}}
        
        & \multirow{2}{*}{\begin{tabular}[c]{@{}c@{}} CodeSe- \\ archNet \end{tabular}}
        % CodeSearchNet
        & % How would you extend the following code to add the features "undo" and "redo" while maintaining a constant time complexity for adding items?
% \vspace{+1em}
\begin{lstlisting}[boxpos=t]
def precedes(self, other):
    """Returns a boolean indicating whether
    this element precedes other element"""
    try:
        ancestor = next(commonancestors(
        AbstractElement, self, other))
    except StopIteration:
        raise Exception("Elements share
        no common ancestor")
    # now we just do a depth first search 
    # and see who comes first
    def callback(e):
        .... Code truncated for brevity
\end{lstlisting}
 & Returns a boolean indicating whether this element precedes the other element & 
        \multirow{2}{*}{\begin{tabular}[c]{@{}c@{}} Document \\  String \end{tabular} } \\
        % Document String\\
        \midrule

        \parbox[t]{3mm}{\multirow{55}{*}{\rotatebox[origin=c]{90}{\textbf{Hybrid Code Query to Hybrid Code Corpus}}}} 
        
        & \multirow{2}{*}{\begin{tabular}[c]{@{}c@{}} StackOver \\ Flow QA \end{tabular}} &  With Rust Regular Expressions, how can I use named capture groups preceding a string? I'm using the sd tool which uses rust regular expressions, and I am trying to use it with a named capture group, however... \textit{<Text ommitted for brevity>} Here is a contrived example to illustrate it:
 \vspace{+1.2em}
 \begin{lstlisting}
 echo `abc' | sd -p `(?P<cg>b)' `$cgB'
 # outputs: ac
 # desired output: abBc
 echo `abc' | sd -p `(?P<cg>b)' `$cg B'
 # outputs as expected: ab Bc
 # however, places a space there
\end{lstlisting}
\vspace{+1.2em}
I've tried \texttt{\$<cg>B, \$cg(B), \$cg0B}, all don't give \texttt{abBc}. I've also checked the rust regex docs however the x flag, and other techniques seem only applicable to the search pattern, not the replace pattern. & We don't need the sd tool to reproduce this behavior. Here it is in pure Rust:
\vspace{+1.2em}
\begin{lstlisting}
let re = regex::Regex::new(r"(?P<n>b)")
                .unwrap();
let before = "abc";
assert_eq!(re.replace_all(before, "$nB"), "ac");
assert_eq!(re.replace_all(before, "${n}B"), "abBc");    
\end{lstlisting}
\vspace{+1.2em}
The brace replacement syntax isn't described in the front documentation but on that of the replace method:

The longest possible name is used. e.g., \texttt{\$1a} looks up the capture
group named \texttt{1a} and not the capture group at index 1. To exert more
precise control over the name, use braces, e.g., \texttt{\$\{1\}a}.
\textit{<Text omitted for brevity>} 
% In short, unless there's a character that can't be part of a name just after in the replacement pattern, you should always put the group name between braces.
% A: For the rust regexes via sd via bash use case in the question:
% \texttt{echo `abc' | sd -p `(?P<cg>b)' `\$\{cg\}B'}
        & \multirow{2}{*}{\begin{tabular}[c]{@{}c@{}} Code Snippet \\ mixed with text \end{tabular}} \\
        % & Code Snippets mixed with text \\
        \cmidrule{2-5}
        &  \multirow{2}{*}{\begin{tabular}[c]{@{}c@{}} CodeFeed \\ Back-ST \end{tabular} }
        & \textcolor{blue}{\textbf{user:}} How would you extend the following code to add the features ``undo" and ``redo" while maintaining a constant time complexity for adding items? 
\vspace{+1.2em}
\begin{lstlisting}[boxpos=t]
def add_item(item, list):
    list.append(item)
def undo(list):
    if len(list) > 0:
        list.pop()
def redo(undone_list, list):
    if len(undone_list) > 0:
        item = undone_list.pop()
        list.append(item)
\end{lstlisting}
        & \textcolor{red}{\textbf{assistant:}} To add the features "undo" and "redo" while maintaining a constant time complexity for adding items, you can use ...
% two additional lists to keep track of the undone items and the redone items. 
Here's an extended version of the code:
\vspace{+1.2em}
\begin{lstlisting}
def add_item(item, item_list, undone_list, redone_list):
    item_list.append(item)
    redone_list.clear()  # Clear the  
        # redone_list as new item is added
def undo(item_list, undone_list, redone_list):
    if len(item_list) > 0:
        .... Code omitted for brevity
\end{lstlisting}\vspace{+1.1em}
In this version, we add three new parameters to the functions: `undone\_list'...
% and `redone\_list`. `undone\_list` keeps track of the undone items, and `redone\_list` keeps track of the redone items. 
\textit{<Text omitted for brevity>}
        & \multirow{2}{*}{\begin{tabular}[c]{@{}c@{}} Code Snippet \\ or Functions \\ mixed with text \end{tabular}} \\
        % & Code Snippets / Functions mixed with Text \\
        \cmidrule{2-5}
        & \multirow{2}{*}{\begin{tabular}[c]{@{}c@{}} CodeFeed \\ Back-MT \end{tabular}}  & \textcolor{blue}{\textbf{user:}} Convert a given 2D point to the polar representation. The point has coordinates (x,y), where x and y are integers between -1000 and 1000, inclusive.
Additional Requirement: ...\textit{<Text omitted for brevity>}
\vspace{+0.6em}

\textcolor{red}{\textbf{assistant:}} To convert a 2D point to polar representation, we need to calculate the polar coordinates (r, theta) using the given Cartesian coordinates (x, y).

\vspace{+0.6em}

We can use the arctangent function to calculate the value of theta. \textit{<Text omitted for brevity>} Now, let's write the code to convert the given 2D point to polar representation using the atan2 function in Python:
\vspace{+1.2em}
\begin{lstlisting}
import math
def convert_to_polar(x, y):
    r = math.sqrt(x*x + y*y) # calculating the distance from the origin
    theta = math.atan2(y, x) # calculating the angle in radians using atan2 function
    return (r, theta)
... Code truncated for brevity
\end{lstlisting}
\vspace{+1.2em}
\textcolor{blue}{\textbf{user:}} \textit{<Text omitted for brevity>},
\vspace{+0.6em}

\textcolor{red}{\textbf{assistant:}} \textit{<Text omitted for brevity>}, 
\vspace{+0.6em}

\textcolor{blue}{\textbf{user:}} Please generate the code to meet the original requirements without using arithmetic operations. Clarifying comments and documentation will also be appreciated.
        &  \textcolor{red}{\textbf{assistant:}} I apologize for any misunderstanding, but converting from Cartesian coordinates to polar coordinates inherently requires mathematical operations -- notably, addition, multiplication, and trigonometric operations. These are defined by the mathematical formulas for converting between the two coordinate systems: $$r = sqrt(x^2 + y^2), theta = atan2(y, x)$$

These formulas can't be implemented without using arithmetic operations, due to their inherent mathematical nature. As per your requirement of not using any arithmetic operations to solve this problem, it's essential to understand that it would be impossible to convert to polar coordinates, as the problem mainly depends on these arithmetic operations. Again, I emphasize that this kind of computation relies heavily on arithmetic operations, and it is a significant constraint that would make finding a software or coding solution impossible. However, if there's any related issue or a different problem you need assistance with, I would be more than happy to assist!"
        & \multirow{2}{*}{\begin{tabular}[c]{@{}c@{}} Code Snippet \\ or Functions \\ mixed with text \end{tabular}} \\
        % & Code Snippets / Functions mixed with Text\\
    
    \bottomrule
    \end{tabular}}
    
    \caption{Examples of queries and relevant corpora for 4 datasets related to Code-to-Text and Hybrid Code retrieval (i.e. \emph{CodeSearchNet}, \emph{StackOverflow QA}, \emph{CodeFeedback-ST} and \emph{CodeFeedback-MT})  in \name. For brevity, we omit or truncate some portion of the query and corpus}
    \label{tab:examples2}
\end{table*}
% \begin{figure}[h]
%     \begin{minipage}{.4\linewidth}
%         \includegraphics[width=1\textwidth]{figures/similarity.pdf}
%         % \label{fig:coir_heatmap}
%     \end{minipage} 
%     \vspace{-1em}
%     \caption{Embedding similarity between different datasets in \name using BGE-Base encoder}
%     \label{fig:coir_emb_distribution}
% \end{figure}

\subsection{Text-to-Code Retrieval Datasets}
\textbf{APPS}~\cite{Hendrycks-etal-apps}. The original APPS dataset is a code generation dataset derived from programming problems shared on open-access sites frequented by programmers, including Codewars, AtCoder, Kattis, and Codeforces, where each example consists of a question description and its corresponding code solution. 
To adapt the original APPS dataset for retrieval, we use the original problem descriptions as the query to retrieve from a corpus of all code solutions. We retain the original dataset's train-test split and remove examples that do not have a corresponding code solution, resulting in a total of 5,000 samples for the training set and 3,765 samples for the test set. 

\textbf{CoSQA}~\cite{huang-etal-2021-cosqa}. The CosQA dataset comprises 20,604 human-annotated labels for pairs of natural language web queries and corresponding code snippets. We retain the original train/dev/test dataset splits of 19,604/500/500 and utilize the natural language web queries as queries to retrieve from a corpus of all code snippets as intended in CoSQA.
% Additionally, the source code encompasses baseline methods along with the proposed contrastive learning method, termed CoCLR, which is designed to improve query-code matching. 
% We utilize web queries as textual input to retrieve corresponding code snippets.

\textbf{Synthetic Text2Sql}~\cite{gretel-synthetic-text-to-sql-2024}. The Synthetic Text2Sql dataset is a comprehensive collection of high-quality synthetic Text-to-SQL samples, meticulously designed and generated using Gretel Navigator \footnote{https://docs.gretel.ai/}. Each example consists of the following: (1) a problem description which can be resolved using SQL (2) information on the schema of relevant tables used (3) the corresponding SQL code solution (4) meta-data revolving problem described, for instance the type of domain or industry the problem falls under (e.g. healthcare, aerospace etc.) and the nature of the task at hand (e.g. reporting, analytics, dashboarding etc.). Here we use the natural language question description as the query to retrieve from a corpus of corresponding SQL code solutions. We follow the train-test split provided by Gretel where the 105,851 queries are divided into 100,000 queries for train and the remaining 5,851 for test. 
% statements.

\subsection{Code-to-Code Retrieval Datasets}
\textbf{CodeSearchNet-CCR}~\cite{husain2019codesearchnet}. CodeSearchNet-Code Context Retrieval (CCR) is modified from the original CodeSearchNet dataset with 1 million (docstring, code) pairs sourced from open-source repositories hosted on GitHub. For each example present in the dataset, we randomly divide each code function into two code segments where all the initial segment serves as our query and all latter segments forms the corpus to be retrieved using the corresponding initial segment. The length of each query is uniformly and randomly selected to comprise between 40\% and 70\% of the total number of characters in the original code or function. We retain the original train/dev/test split of 905k/41k/53k and also retain the partition by their respective language from CodeSearchNet.

\textbf{CodeTransOcean-DL}~\cite{yan-etal-2023-codetransocean}. CodeTransOcean-DeepLearning (DL) is derived from the DLTrans, a dataset featured in the CodeTransOcean benchmark which focuses on code translation. The original dataset consists of pairs of semantically equivalent deep learning code written using different deep learning libraries namely Tensorflow, Pytorch, PaddlePaddle and MXNet. For each pair of data, we use one of the codes in one of the framework as query to retrieve its counterpart written in a different framework. We follow the original train/dev/test split of 564/72/180 in CodeTransOcean-DL. 
% The original dataset entails translating deep learning code written using the PyTorch library into equivalent deep learning code written using TensorFlow or PaddlePaddle libraries. Each code

\textbf{CodeTransOcean-Contest}~\cite{yan-etal-2023-codetransocean}.
% CodeTransOcean-Contest is similar to CodeTransOcean-DL, as both are subsets of the CodeTransOcean datasets, but focuses on translating one programming language to another. 
Similar to CodeTransOcean-DL, the CodeTransOcean-Contest dataset is derived from the MultilingualTrans dataset in the CodeTransOcean benchmark. The dataset features a collection of code for various problems, for instance binary tree traversal, sorting algorithms, written in various languages featured in Rosetta Code \footnote{https://rosettacode.org/wiki/Rosetta\_Code}, a programming chrestomathy website. In CodeTransOcean-Contest, we focus on retrieving semantically equivalent C++, 
% from its counterpart written in 
Python code pairs as these two languages differs greatly in terms of syntax and language features. We filter the examples in MultilingualTrans that does not contain any C++ nor Python code pairs and we use the Python code in each pair as queries to retrieve its counterpart from all C++ code in the filtered dataset. We retain the dataset split as in MultilingualTrans resulting in a 561/226/221 examples for train/dev/test respectively. 
% We utilize this dataset to retrieve code written in another language using code written in one language.

\subsection{Code-to-Text Retrieval Datasets}
\textbf{CodeSearchNet}~\cite{husain2019codesearchnet}. The CodeSearchNet  is a dataset consisting of 1 million (docstring, code) pairs sourced from open-source repositories hosted on GitHub. It contains code and documentation for several programming languages. Instead of retrieving relevant code from the original description, we reverse the roles of the docstring and code by retrieving the relevant docstring using the code as a query. We follow the original CodeSearchNet train/dev/test split of 905k/41k/53k and also retain the partition by their respective language.

\subsection{Hybrid Code Retrieval Datasets}
\textbf{StackOverflow QA}\footnote{https://www.kaggle.com/datasets/stackoverflow/stacksample/data}.
We modify the original StackOverflow dataset from Kaggle \footnote{https://www.kaggle.com/}, which contains questions posted by users, the corresponding highest voted answer to the user's questions and tags pertaining to the user's questions. Both the question and the answer typically consist of a mix of text and code snippets. Out of a total of 1,048,576 questions, we randomly sample 19,931 questions and their corresponding answers from the StackOverflow QA dataset. We use the questions raised by the users as a query to retrieve the corresponding answer from other correct answers posted by users. We randomly split the data into train/dev/test sets with a 13,951/3,986/1,994 split respectively.  
% The StackOverflow QA dataset is constructed by us from the original StackOverflow dataset. We sample questions and their highest upvoted answers, then split the data into training, validation, and test sets with a ratio of 7:2:1. In the context of CoIR, we use the questions to retrieve their corresponding answers.

\textbf{CodeFeedBack-ST}~\cite{zheng2024opencodeinterpreter}. We utilize data from CodeFeedback-Filtered-Instruction to generate retrieval dataset for CodeFeedBack-ST. CodeFeedback-Filtered-Instruction is a collection of instruction-tuning datasets where a language model takes cue from user's instruction in order to generate code to fulfil the user's instructions. The dataset consists of pairs of users instruction and the response to be expected from an assistant where both the user's instructions and the assistant's response could contain a mixture of text and code snippets. We use the user's instructions as query to retrieve the corresponding relevant expected assistant's reply. Of the 156,526 examples present, we split the train and test dataset into 125,221 and 31,307 train and test examples respectively.

\textbf{CodeFeedBack-MT}~\cite{zheng2024opencodeinterpreter}. We utilize examples from Code-Feedback dataset to generate retrieval dataset for CodeFeedBack-Multi Turn (MT). Code-Feedback is a collection of instruction-tuning datasets involving simulated multi-turn dialogues between 2 LLMs; one plays the role of a user and the other plays the role of assistant. To ensure that that the generated replies of the assistant are of high quality, the LLM playing the role of assistant is aided by code compilers alongside multiple generation attempts to arrive at a desirable response. Each example comprises a sequence of exchanges alternating between the user and assistant, starting with the user and ending with the assistant. To construct CodeFeedBack-MT, we split each sequence into 2 portions, the first portion consists of the initial dialogue history, prior to the last reply by the assistant, and second part is made up of the final reply by the assistant. We use the initial portion of the dialogue as query to retrieve the corresponding reply by the assistant. We split the dataset into 53,106 and 13,277 for train and test respectively. 

\section{Metric}
\subsection{NDCG Metric}
Normalized Discounted Cumulative Gain (NDCG) is a popular metric used to evaluate the quality of rankings, particularly in information retrieval and recommender systems. It measures the usefulness, or \emph{gain}, of an relevant item based on its position in the result list, discounted logarithmically by the position.

The Discounted Cumulative Gain (DCG) is the sum of the gains of relevant items, discounted logarithmically by their positions in the ranking. The normalized version, NDCG, compares the DCG of the ranked list to the DCG of the ideal ranking. This normalization ensures that the score is within the range [0, 1].

\textbf{DCG Calculation:}

The DCG at position \( p \) is calculated as:

$$
\text{DCG}_p = \sum_{i=1}^{p} \frac{2^{rel_i} - 1}{\log_2(i + 1)}
$$

where \( rel_i \) is the relevance score of the item at position \( i \).

\textbf{IDCG (Ideal DCG) Calculation:}

The Ideal DCG (IDCG) is the DCG of the ideal ranking. This is calculated by sorting all items by their relevance scores in descending order and then computing the DCG using the same formula.

$$
\text{IDCG}_p = \sum_{i=1}^{p} \frac{2^{rel^{\ast}_i} - 1}{\log_2(i + 1)}
$$

where \( rel^{\ast}_i \) is the relevance score of the item at position \( i \) in the ideal ranking.

\textbf{NDCG Calculation:}

The NDCG at position \( p \) is the ratio of the DCG at position \( p \) to the IDCG at position \( p \):

$$
\text{NDCG}_p = \frac{\text{DCG}_p}{\text{IDCG}_p}
$$

This normalization ensures that NDCG values are bounded between 0 and 1, making comparisons across different queries or datasets meaningful.

By evaluating the NDCG, we can get a sense of how well our ranking system is performing relative to the best possible ranking. This is particularly useful when the relevance varies across items and lower-ranked items are less likely to be seen and thus should contribute less to the overall evaluation metric.

\subsection{Weighted Jaccard Similarity}\label{sect:weighted_jaccard_similarity}

The weighted Jaccard similarity $J(S,T)$ \cite{ioffe2010improved} is calculated as the unique word overlap for all words present in both the datasets. We define the normalized frequency for an unique word $k$ in a dataset to be the frequency of word $k$ divided over the sum of frequencies of all words in the dataset.

Let ${S_k}$ be the normalized frequency of word $k$ in the source dataset $S$ and ${T_k}$ for the target dataset $T$ respectively. The weighted Jaccard similarity between $S$ and $T$ is defined as:

\[ J(S, T) = \frac{\sum_{k}\min(S_k, T_k)}{\sum_{k}\max(S_k, T_k)}\] 

where the sum is over all unique words $k$ present in datasets $S$ and $T$.

\section{Implementation Details} \label{appendix:imple_details}

\subsection{Implementation Environments} \label{appendix:setup}
 The experimental setup for performance and efficiency evaluation was conducted utilizing a Tesla V100 GPU  with 32GB of memory and Intel Titan CPU and CUDA 11.2. The versions of the software utilized in this study are PyTorch 2.0.1 and Transformers 4.38.1. We utilized the Faiss IndexFlat\cite{johnson2019billion} to store vector embedding index.

% \subsection{Training Details} \label{appendix:train_details}
% \textcolor{red}{specify all the training details (e.g., data splits, hyperparameters, how they were chosen)}

\subsection{Retriever Resources} \label{appendix:retrievers}
\begin{table*}[htbp]
    \small
    % \resizebox{\textwidth}{!}{
    \begin{tabular}{ l | l }
        \toprule
        \multicolumn{1}{l|}{\textbf{Model}} &
        \multicolumn{1}{c}{\textbf{Public Model Checkpoints (Link)}} \\ \midrule
        Contriever & \url{https://huggingface.co/facebook/contriever-msmarco} \\
        E5-base & \url{https://huggingface.co/intfloat/e5-base-v2} \\
        BGE-Base & \url{https://huggingface.co/BAAI/bge-base-en-v1.5} \\
        GTE-Base & \url{https://huggingface.co/Alibaba-NLP/gte-base-en-v1.5} \\
        UniXcoder & \url{https://huggingface.co/microsoft/unixcoder-base} \\
        BGE-M3 & \url{https://huggingface.co/BAAI/bge-m3} \\
        E5-Mistral & \url{https://huggingface.co/intfloat/e5-mistral-7b-instruct} \\
        OpenAI-Ada-002 & \url{https://openai.com/} \\
        Voyage-Code-002 & \url{https://www.voyageai.com/} \\
        \bottomrule
    \end{tabular}
    % }
    \caption{Publicly available model links used for evaluation in \textbf{\name}.}
    \label{tab:model_links}
\end{table*}
The huggingface models or API used of retrieval model used in \name is listed in Table~\ref{tab:model_links}.
% \textcolor{red}{Here, the huggingface models or API used.}

\subsection{Dataset Resources} \label{appendix:dataset_url}
\begin{table*}[htbp]
    \small
    \resizebox{\textwidth}{!}{
    \begin{tabular}{ l | l }
        \toprule
        \multicolumn{1}{l|}{\textbf{Corpus}} &
        \multicolumn{1}{|c}{\textbf{Website (Link)}} \\
        \midrule
        APPS & \url{https://huggingface.co/datasets/codeparrot/apps} \\
        CoSQA & \url{https://github.com/microsoft/CodeXGLUE/tree/main/Text-Code/NL-code-search-WebQuery} \\
        Synthetic Text2sql & \url{https://huggingface.co/datasets/gretelai/synthetic_text_to_sql} \\
        CodeSearchNet & \url{https://huggingface.co/datasets/code-search-net/code_search_net} \\
        CodeTransOcean & \url{https://huggingface.co/datasets/WeixiangYan/CodeTransOcean} \\
        Code-FeedBack & \url{https://huggingface.co/datasets/m-a-p/Code-Feedback} \\
        CodeFeedBack-Filtered-Instruction & \url{https://huggingface.co/datasets/m-a-p/CodeFeedback-Filtered-Instruction}\\
        \bottomrule
    \end{tabular}
    }
    \caption{Corpus Name and Link used for datasets in \textbf{\name}.}
    \label{tab:corpus_links}
\end{table*}
The urls of dataset resources used in \name is listed in Table~\ref{tab:corpus_links}.
\section{Dataset Explanation of Reality} 
 It is important to note that the current settings align with prevalent code retrieval benchmark methodologies in the field, such as CodeSearchNet and CoSQA, where the retrieved documents are considered the final answers to queries. Additionally, we have considered scenarios where retrieval serves as a stepping stone rather than just an endpoint. Specifically, our code-to-text and code-to-code tasks are designed to evaluate the model's ability to retrieve relevant text and code snippets that supplement and enhance the final answer.

Furthermore, we have developed a new dataset from Stack Overflow. This dataset is carefully curated and filtered from the original Stack Overflow dump. In this dataset, the provided answers are not always the final solution to the queries. Instead, they may represent a strategy for solving the problem or serve as a component of the solution, thus reflecting the nuanced and iterative nature of real-world code generation and retrieval tasks.
\section{Clarification of Innovation}
We are not merely integrating existing datasets; rather, we are introducing new datasets and tasks. CodeSearchNet-CCR and StackOverflowQA are datasets we created ourselves. StackOverflowQA was extracted from the original StackOverflow dump, and we introduced this dataset to ensure that the simulated tasks closely mirror real-world scenarios, such as searching for answers when encountering code issues.

Additionally, we introduced a new task, Code Context Retrieval, which involves retrieving the most relevant code segment that completes a given initial segment of code. This task is critical for code completion purposes, and the CodeSearchNet-CCR dataset was specifically created for this task.

Previously, only CosQA and CodeSearchNet were straightforwardly usable as code retrieval benchmarks. We have supplemented these with eight additional datasets, which we meticulously cleaned, manually filtered, and standardized. This process was quite labor-intensive.
\label{sec:appendix}

\end{document}